\documentclass[amssymb,amsmath,aps,prb,showpacs,twocolumn]{revtex4}
\usepackage{times}
\usepackage{graphicx}
\usepackage{bm}
\usepackage{latexsym}
\usepackage{dcolumn} 

\begin{document}



\title{Electromechanical coupling in free-standing AlGaN/GaN planar structures}

\author{B. Jogai}
\affiliation{Air Force Research Laboratory, Materials and
Manufacturing Directorate, Wright-Patterson Air Force Base, Ohio
45433} \affiliation{Semiconductor Research Center, Wright State
University, Dayton, Ohio 45435}

\author{J. D. Albrecht}
\affiliation{Air Force Research Laboratory, Wright-Patterson Air
Force
   Base, Ohio 45433}
\author{E. Pan}
\affiliation{Department of Civil Engineering, The University of
Akron, Akron, Ohio 44325}


\begin{abstract}
  The strain and electric fields present in free-standing AlGaN/GaN
  slabs are examined theoretically within the framework of
  fully-coupled continuum elastic and dielectric models.  Simultaneous
  solutions for the electric field and strain components are obtained
  by minimizing the electric enthalpy. We apply constraints
  appropriate to pseudomorphic semiconductor epitaxial layers and
  obtain closed-form analytic expressions that take into account the
  wurtzite crystal anisotropy. It is shown that in the absence of free
  charges, the calculated strain and electric fields are substantially
  differently from those obtained using the standard model without
  electromechanical coupling.  It is also shown, however, that when a
  two-dimensional electron gas is present at the AlGaN/GaN interface,
  a condition that is the basis for heterojunction field-effect
  transistors, the electromechanical coupling is screened and the
  decoupled model is once again a good approximation. Specific cases
  of these calculations corresponding to transistor and superlattice
  structures are discussed.
\end{abstract}
\maketitle

\section{Introduction}
\label{sec:intro} In previous analyses of the strain in epitaxial
layers of AlGaN/GaN, the electrical and mechanical properties of
the crystal have been treated as though they are decoupled. Linear
elastic theory is assumed to hold, and Hooke's law is invoked to
obtain the relation between the in-plane and axial components of
the strain tensor.  The standard model decouples the strain tensor
from the electric field and enables the separation of the electric
field and electronic eigenstate calculations from calculations of
the strain field. Historically, there is a solid tradition of
using this separability when studying heterostructures of GaAs and
associated alloys where the approximation that the strain field is
negligibly affected by the electric field is valid.\cite{smith}

Generally, it is known from thermodynamics\cite{Landau2} that the
electrical and mechanical properties of piezoelectric materials
are coupled and a simultaneous treatment is called for when the
electromechanical coupling is significant. This is the case when
the piezoelectric response is large, as it is in AlGaN, especially
for high Al fractions. For example, large corrections of the
electrostatic and elastic properties have been predicted for
AlGaN/GaN transistor structures\cite{jogainew} and for
free-standing plates of AlN\cite{Pan1} when a fully-coupled model
is used instead of a standard (uncoupled) one.

In this paper, we calculate the strain field in a free-standing
AlGaN/GaN slab using a fully-coupled model.  A free-standing bi-layer
slab is chosen as the model structure on which to develop the theory
as it can serve as a building block for other, more complicated,
structures.  For instance, it can be used as a building block for
superlattices and quantum wells.  In addition, a heterojunction
field-effect transistor (HFET) is just a special case of a bi-layer
slab in the limit that the GaN layer is much thicker than the AlGaN
layer.  It will be seen that the fully-coupled strain and electric
fields for an HFET can be obtained in this asymptotic limit.

Although the present model is still within the framework of continuum
mechanics, it goes beyond the standard continuum elastic theory used
in typical strain calculations in semiconductor materials.  We obtain
the total electric enthalpy for the bi-layer slab and minimize it
subject to certain constraints to find both the strain and electric
fields.  Two constraints are used.  One is that the two layers must
share a common $c$-plane lattice constant after strain.  This lattice
constant is unknown at the outset and is deduced only after
minimization.  The other constraint is set up by the electrostatic
potential, and hence the electric field, being forced to satisfy the
Poisson equation subject to the boundary conditions at the surface and
the common interface.  Both the spontaneous and piezoelectric
polarizations are included.  The effect on the strain and electric
fields of a two-dimensional electron gas (2DEG) at the AlGaN/GaN
interface, a situation that is essential for channel conduction in the
HFET, is also investigated. The model produces expressions for the
strain tensor and the electric field in the two layers in closed form.

This paper is organized as follows: In Sec.\ \ref{sec:model}, the
electric enthalpy for the AlGaN/GaN slab is minimized subject to the
constraints derived from the pseudomorphic strain condition and from
the Poisson equation.  The strain tensor and the electric field are
derived in closed form.  In Sec.\ \ref{sec:results}, the calculated
results from the fully-coupled model are contrasted with those from
the uncoupled model.  The effect on the strain and electric fields of
screening from the 2DEG is discussed.  The results are summarized in
Sec.\ \ref{sec:summary}.

\section{Model description}
\label{sec:model}
In the standard strain theory for semiconductor materials, the formal
way to calculate the strain field in a generalized strain problem is
to minimize the Helmholtz free energy\cite{Nye,Landau}
\begin{equation}
  \label{eq:Fdef}
  F = \frac{1}{2} C_{ijkl} \gamma_{ij} \gamma_{kl} ,
\end{equation}
in which $ C_{ijkl} $ is the fourth-ranked elastic stiffness
tensor, $ \gamma_{ij} $ is the strain tensor, and the indices $i$,
$j$, $k$, and $l$ run over the Cartesian coordinates $x$, $y$, and
$z$. Summation over repeated indices is implied throughout. The
minimization will, of course, be subject to certain constraints
brought about by the boundary conditions at the surfaces and
between adjacent materials, as well as any external forces.

In a wide range of semiconductor problems, however, one is
concerned with calculating the strain in epitaxial layers grown,
at least in principle, pseudomorphically on a thick buffer layer.
Typically, the thick layer belongs to, or at least closely
resembles, the same crystallographic point group as the epilayers.
In this case, a minimization of Eq.\ (\ref{eq:Fdef}) is
unnecessary, since we can take full advantage of the boundary
condition for a free surface which requires that the components of
the stress tensor parallel to an outward normal to the surface be
zero.  For a surface normal to the $z$-axis, this means that $
\sigma_{iz} = 0 $, where $ \sigma_{ij} $ is the stress tensor.  In
addition, since the lateral extent of the layers is far greater
than their thickness, a one-dimensional (1-D) approximation can be
used, wherein $ \gamma_{ij}$$=$$0$ for $i$$\neq$$j$.  This
condition is equivalent to ignoring the effects of bowing on the
local lattice displacement. Under these conditions, the result
\begin{equation}
  \label{eq:poisratio}
  \sigma_{zz} = 0 = C_{xxzz} ( \gamma_{xx} + \gamma_{yy} ) + C_{zzzz}
  \gamma_{zz} ,
\end{equation}
is obtained for wurtzite epilayers oriented in the [0001]
direction. Here the only unknown is $ \gamma_{zz} $, because a
standard approach is to assume that the buffer layer is unstrained
and that the epilayer assumes the $c$-plane lattice constant of
the buffer, effectively fixing $\gamma_{xx}$ and $\gamma_{yy}$.
Clearly, when $\gamma_{xx}$$=$$\gamma_{yy}$$=$$(a$$-$$a_o)/a_o$,
Eq.\ \ref{eq:poisratio} yields the usual result for the Poisson
effect in the case of biaxial strain in an epilayer which is
$\gamma_{zz}/\gamma_{xx}$$=$$-2C_{xxzz}/C_{zzzz}$.

\begin{figure}
  \begin{center}
    \includegraphics[width=3.2in]{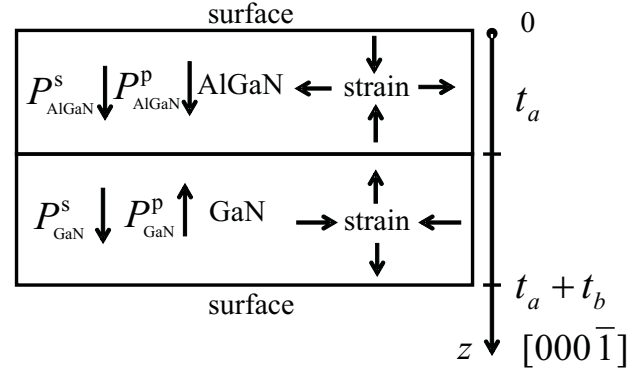}
  \end{center}
  \caption{Cross-section of a model cation-faced AlGaN/GaN bi-layer
    slab showing the direction of the piezoelectric $ P^\textrm{p} $
    and spontaneous $ P^\textrm{s} $ polarization vectors in relation
    to the $z$-axis. Both layers in the slab are under strain as
    shown.  $t_a$ and $t_b$ are the thicknesses of the AlGaN and GaN
    layers, respectively.}
  \label{fig:schematic}
\end{figure}

The situation for the free-standing bi-layer slab depicted in Fig.\
\ref{fig:schematic} in which the two layers are of comparable
thicknesses is more complicated, since none of the strain components
is known beforehand.  To obtain the strain, at least within the
standard model, the total mechanical strain energy for the slab must
be minimized with respect to $ \gamma_{ij} $ subject to the
constraints along the interface.  There is a further complication in
piezoelectric materials such as AlGaN and GaN wherein the electrical
and mechanical properties are coupled through the piezoelectric
coefficient tensor.  Equation (\ref{eq:Fdef}) is no longer a suitable
energy functional for calculating the strain field, since it does not
include the electromechanical coupling.  Instead, we begin the
derivation with the electric enthalpy $H$ given by\cite{ANSI}
\begin{equation}
  \label{eq:Hdef}
 H = U - \textbf{E}\cdot\textbf{D} ,
\end{equation}
where $\textbf{E}$ and $\textbf{D}$ are the electric field and
electric displacement, respectively, and $ U $ is the total
internal energy given by
\begin{equation}
  \label{eq:U}
  U = \frac{1}{2} C_{ijkl}
\gamma_{ij} \gamma_{kl} + \frac{1}{2} \varepsilon_{ij} E_i E_j,
\end{equation}
in which $ \varepsilon_{ij} $ is the tensor form of the electric
permittivity.  The electric displacement field components are
given by
\begin{equation}
  \label{eq:Ddef}
  D_i = e_{ijk} \gamma_{jk} +\varepsilon_{ij} E_j+ P^\textrm{s}_i,
\end{equation}
in which $ e_{ijk} $ is the piezoelectric coefficient tensor and $
P^\textrm{s}_i $ is the spontaneous polarization.\cite{Bernardini} The
first term in Eq.\ (\ref{eq:Ddef}) is recognized as the piezoelectric
polarization.  The spontaneous polarization is in the direction $
\textrm{N} \to \textrm{Ga} $ along the $c$-axis bond.  After
substitution into Eq.\ (\ref{eq:Hdef}), the enthalpy in its final form
is
\begin{equation}
  \label{eq:Hfinal}
  H = \frac{1}{2} C_{ijkl} \gamma_{ij} \gamma_{kl} -  e_{ijk}
  E_i \gamma_{jk} - \frac{1}{2} \varepsilon_{ij} E_i
  E_j - E_i P^\textrm{s}_i .
\end{equation}
This expression includes the full electromechanical coupling as
well as the spontaneous polarization.  Differentiating the
enthalpy with respect to the strain tensor gives the fully-coupled
equation of state
\begin{equation}
  \label{eq:state}
  \sigma_{ij} = C_{ijkl} \gamma_{kl} - e_{kij} E_k ,
\end{equation}
for piezoelectric materials.  To obtain the strain and electric fields
within a fully-coupled model, Eq.\ (\ref{eq:Hfinal}) must be minimized
with respect to both $ \gamma_{ij} $ and $ E_i $ subject to the
constraints to be discussed shortly for the bi-layer slab of Fig.\
\ref{fig:schematic}.

Using the Voigt notation\cite{Nye} and expanding Eq.\
(\ref{eq:Hfinal}), the enthalpy for a wurtzite crystal with the
[0001] axis as the principal axis can be written as
\begin{eqnarray}
  \label{eq:voigt}
H = \frac{1}{2} C_{11} ( \gamma_{xx}^2 + \gamma_{yy}^2 ) + \frac{1}{2}
C_{33} \gamma_{zz}^2 + C_{12} \gamma_{xx} \gamma_{yy} + \nonumber \\
C_{13} \gamma_{zz} ( \gamma_{xx} +
\gamma_{yy} ) + 2 C_{44} ( \gamma_{xz}^2 +
\gamma_{yz}^2 ) +  \nonumber \\ ( C_{11} - C_{12} ) \gamma_{xy}^2
- E_z [ e_{31} ( \gamma_{xx} + \gamma_{yy} ) + e_{33} \gamma_{zz} ] -
\nonumber \\ e_{15} ( E_x \gamma_{xz} + E_y \gamma_{yz} )
- E_z P^\textrm{s} - \frac{1}{2} \epsilon ( E_x^2 + E_y^2 + E_z^2 ) ,
\end{eqnarray}
where $ \epsilon $ is the electric permittivity assumed here, for
simplicity, to be a scalar.  The enthalpy for the slab is given by
\begin{equation}
  \label{eq:Hslab}
  H^\textrm{slab} = \frac{ t_a  H^\textrm{a} + t_b H^\textrm{b} }{ t_a
  + t_b } ,
\end{equation}
where $ t_a $ and $ t_b $ are the thicknesses of the two layers
and ``a'' denotes the AlGaN layer and ``b'' the GaN layer.  It is
noted that the elastic and piezoelectric coefficients and the
electric permittivity can be substantially different in the two
layers. These variations are taken into account in Eq.\
(\ref{eq:Hslab}). Because of the rotational symmetry of the slab,
$ \gamma_{yy} = \gamma_{xx} $ and $ \gamma_{yz} = \gamma_{xz} $ in
each layer. The shear terms in the strain tensor, $ \gamma_{ij} $
for $ i \ne j $, turn out to be zero because of the assumption
that the strain is piecewise homogeneous.  A more realistic
assumption would be to expect the strain to diminish away from the
interface, resulting in a non-zero $ \gamma_{xz} $ that would vary
with position along the $c$ axis.  This situation would be
manifested in a bowing of the slab.  Such inhomogeneous strains,
however, would require a three-dimensional (3-D) numerical
calculation that is beyond the scope of the present work.

One of the minimization constraints on Eq.\ (\ref{eq:Hslab}) is that
the two layers share the same $c$-plane lattice constant which is
unknown at the outset.  This condition is expressed as
\begin{equation}
  \label{eq:face}
  (1 + \gamma_{xx}^\textrm{a} ) a_a = (1 + \gamma_{xx}^\textrm{b} ) a_b ,
\end{equation}
where $a_a$ and $a_b$ are the unstrained $c$-plane lattice of the
AlGaN and GaN layers, respectively.  Equation (\ref{eq:face})
presumes ideal growth conditions.  Partial relaxation due to
dislocations will be the subject of future investigations.

The other constraint is the relationship between the electric and
strain fields.  This relationship is established in the present work
by solving the 1-D Poisson equation given by
\begin{equation}
  \label{eq:pois1D}
  \frac{\partial}{\partial z} \epsilon \frac{\partial \phi }{\partial
  z} = \frac{\partial }{ \partial z } ( P^\textrm{s} + 2
  e_{31} \gamma_{xx} + e_{33} \gamma_{zz} ) + e_0 n_\textrm{2D}
  \delta( z - t_a ) ,
\end{equation}
where $ \phi $ is the electrostatic potential and $ n_\textrm{2D}
$ is the sheet concentration of an ideal 2DEG modeled as a
$\delta$-function at the AlGaN/GaN interface.  Equation
(\ref{eq:pois1D}) is solved analytically for $ \phi $ subject to
the boundary conditions $ \phi = 0 $ at $ z = 0 $ and $ z = (t_a +
t_b ) $ and also to the continuity of $ \phi $ and the electric
displacement across the interface.  The general solution of Eq.\
(\ref{eq:pois1D}) in each layer is given by
\begin{equation}
  \label{eq:gensol}
  \phi = \frac{P^\textrm{s}}{\epsilon} z +\frac{2 e_{31} \gamma_{xx} +
  e_{33} \gamma_{zz}}{\epsilon} z + \frac{A}{\epsilon} z + B ,
\end{equation}
where $A$ and $B$ are unknown constants.  It is evident that there are
four unknowns, two in each layer.  The $B$'s are eliminated by
enforcing the boundary conditions $ \phi = 0 $ at $ z = t_a $ and $ z
= t_a + t_b $.  This condition at the two surfaces presumes that the
polarization charge of the materials is terminated by external charges
and that the applied bias voltage is zero.  Further, it ensures that
the electric field in the free space above and below the slab is zero,
a reasonable and physically plausible result.  The relationship
between the two $A$'s is established from the continuity of the
electric displacement and is obtained by integrating
Eq.\ (\ref{eq:pois1D}) across the interface (see Fig.\
\ref{fig:schematic}) as follows:
\begin{equation}
  \label{eq:Dcontinue}
  \left . \epsilon \frac{\partial \phi}{\partial z} \right
  |^{t_a^+}_{t_a^-} = \left . P^\textrm{s} \right |^{t_a^+}_{t_a^-} +
  \left . (2 e_{31} \gamma_{xx} + e_{33} \gamma_{zz}) \right
  |^{t_a^+}_{t_a^-} + e_0 n_\textrm{2D} .
\end{equation}
From Eqs.\ (\ref{eq:gensol}) and (\ref{eq:Dcontinue}), the relation
between the $A$'s in the respective layers is given by
\begin{equation}
  \label{eq:A1A2}
  A^\textrm{GaN} = A^\textrm{AlGaN} + e_0 n_\textrm{2D} .
\end{equation}
From Eq.\ (\ref{eq:A1A2}) and from the continuity of $\phi$ at $ z =
t_a $, we can solve for all of the remaining unknowns.

With the electrostatic potential now known in terms of the, as yet,
undetermined strain tensor, the electric fields in the two layers are
then given by
\begin{equation}
  \label{eq:Eza}
  E_z^\textrm{a} = t_b \frac{P^\textrm{net} - 2  e_{31}^\textrm{a}
  \gamma_{xx}^\textrm{a} +
  2 e_{31}^\textrm{b} \gamma_{xx}^\textrm{b} - e_{33}^\textrm{a}
  \gamma_{zz}^\textrm{a} + e_{33}^\textrm{b}
  \gamma_{zz}^\textrm{b}}{t_a \epsilon_b +
  t_b \epsilon_a } ,
\end{equation}
and
\begin{equation}
  \label{eq:Ezb}
  E_z^\textrm{b} = - t_a \frac{P^\textrm{net} - 2  e_{31}^\textrm{a}
  \gamma_{xx}^\textrm{a} +
  2 e_{31}^\textrm{b} \gamma_{xx}^\textrm{b} - e_{33}^\textrm{a}
  \gamma_{zz}^\textrm{a} + e_{33}^\textrm{b}
  \gamma_{zz}^\textrm{b}}{t_a \epsilon_b +
  t_b \epsilon_a } ,
\end{equation}
where $ P^\textrm{net} = P^\textrm{s(b)} - P^\textrm{s(a)} + e_0
n_\textrm{2D} $.  Equation (\ref{eq:Hslab}) can be readily minimized
with respect to $ \gamma_{ij} $ and $ E_z $ subject to the constraints
expressed in Eqs.\ (\ref{eq:face}), (\ref{eq:Eza}), and (\ref{eq:Ezb})
using commercial symbolic-mathematical programs.  For convenience, we
have also formulated the problem in the form of a matrix equation of
the form
\begin{equation}
  \label{eq:matrix}
\left (
  \begin{matrix}
    A_{11} & A_{12} & A_{13} \\
    A_{21} & A_{22} & A_{23} \\
    A_{31} & A_{32} & A_{33}
  \end{matrix}
  \right )
  \left (
  \begin{matrix}
    \gamma_{xx}^\textrm{a} \\
    \gamma_{zz}^\textrm{a} \\
    \gamma_{zz}^\textrm{b}
  \end{matrix}
  \right ) =
  \left (
  \begin{matrix}
    B_1 \\
    B_2 \\
    B_3
  \end{matrix}
  \right ) ,
\end{equation}
in which the matrix elements are obtained by differentiating the
energy functional with respect to the unknown strain components.
The constraints are already built into Eq.\ (\ref{eq:matrix}) by
eliminating $ \gamma_{xx}^\textrm{b} $, $ E_z^\textrm{a} $, and $
E_z^\textrm{b} $ from Eq.\ (\ref{eq:Hslab}).  Equation
(\ref{eq:matrix}) can be solved either symbolically or
numerically. We have provided the symbolic solution as an
auxiliary item.\cite{EPAPS} Appendix \ref{app:matrix} lists the
matrix elements.

The fully-coupled model described above reproduces the strain
tensor from the standard model in the limit that the piezoelectric
stress tensor is set to zero.  The results are as follows:
\begin{equation}
  \label{eq:exxa_std}
 \left . \gamma_{xx}^\textrm{a} \right |_{e_{ijk} \rightarrow 0 } =
  \frac{t_b a_a (a_b - a_a) C_{33}^\textrm{a}}{t_b
  a_a^2 C_{33}^\textrm{a} + t_a a_b^2 C_{33}^\textrm{b} R}
\end{equation}
and
\begin{equation}
  \label{eq:ezza_std}
  \left . \gamma_{zz}^\textrm{a} \right |_{e_{ijk} \rightarrow 0 } =
  -2 \frac{t_b a_a (a_b - a_a) C_{13}^\textrm{a}}{t_b a_a^2
  C_{33}^\textrm{a} + t_a a_b^2 C_{33}^\textrm{b} R}
\end{equation}
for the AlGaN layer and
\begin{equation}
  \label{eq:exxb_std}
  \left . \gamma_{xx}^\textrm{b} \right |_{e_{ijk} \rightarrow 0 } = -
  \frac{t_a a_b (a_b - a_a) C_{33}^\textrm{b} R}{t_b a_a^2 C_{33}^\textrm{a}
  + t_a a_b^2 C_{33}^\textrm{b} R}
\end{equation}
and
\begin{equation}
  \label{eq:ezzb_std}
  \left . \gamma_{zz}^\textrm{b} \right |_{e_{ijk} \rightarrow 0 } = 2
  \frac{t_a a_b (a_b - a_a) C_{13}^\textrm{b} R}{t_b a_a^2
  C_{33}^\textrm{a} + t_a a_b^2 C_{33}^\textrm{b} R}
\end{equation}
for the GaN layer where the constant factor $R$ is defined as
\begin{equation}
  \label{eq:r_def}
  R = \frac{( C_{11}^\textrm{a} + C_{12}^\textrm{a} )
  C_{33}^\textrm{a} - 2
  {C_{13}^\textrm{a}}^2}{(C_{11}^\textrm{b} + C_{12}^\textrm{b} )
  C_{33}^\textrm{b} - 2 {C_{13}^\textrm{b}}^2} .
\end{equation}
It is further seen from Eqs.\ (\ref{eq:exxa_std})--(\ref{eq:ezzb_std})
that in the limit $ t_b \gg t_a $, the well-known uncoupled result for
the axial strain in an epitaxial AlGaN layer pseudomorphically
strained on a thick GaN buffer is recovered:
\begin{equation}
  \label{eq:ezza_std_hfet}
  \left . \gamma_{zz}^\textrm{a} \right |_{e_{ijk} \rightarrow 0 , t_b
  \rightarrow \infty } =
  -2 \frac{ C_{13}^\textrm{a}}{ C_{33}^\textrm{a} }
  \left .\gamma_{xx}^\textrm{a} \right |_{t_b \rightarrow \infty} .
\end{equation}
In this limit, the strain in the GaN layer is reduced to zero, at
least ideally, as the thick GaN layer serves as a fixed anchor.  The
uncoupled electric field in the AlGaN layer in this limit then becomes
\begin{multline}
  \label{eq:EAlGaN_std}
  \left . E_z^\textrm{a} \right |_{e_{ijk} \rightarrow 0 , t_b
  \rightarrow \infty} =
  \frac{P^\textrm{net} }{\epsilon_a} \\ - 2 \left (
  \frac{e_{31}^\textrm{a} C_{33}^\textrm{a} - e_{33}^\textrm{a}
  C_{13}^\textrm{a}}{\epsilon_a } \right ) \left . \gamma_{xx}^\textrm{a}
  \right |_{t_b \rightarrow \infty} .
\end{multline}

We can also extract the fully-coupled results for the strain and
electric fields in the limit $ t_b \gg t_a $, i.e. the usual HFET
configuration, and compare them with the uncoupled results.  In this
case, fully-coupled axial strain in the AlGaN layer is given by
\begin{eqnarray}
  \label{eq:ezzcoupled}
  \left .\gamma_{zz}^\textrm{a} \right |_{t_b \rightarrow \infty} = - 2
  \frac{C_{13}^\textrm{a}}{C_{33}^\textrm{a}} \left
  . \gamma_{xx}^\textrm{a} \right |_{t_b \rightarrow \infty} +
   \nonumber \\
  \left ( \frac{2 e_{33}^\textrm{a} (e_{33}^\textrm{a}
  C_{13}^\textrm{a} - e_{31}^\textrm{a}
  C_{33}^\textrm{a})}{C_{33}^\textrm{a}(\epsilon_a C_{33}^\textrm{a} +
  {e_{33}^\textrm{a}}^2) } \right ) \left .\gamma_{xx}^\textrm{a}
  \right |_{t_b \rightarrow \infty}
   + \frac{e_{33}^\textrm{a} P^\textrm{net}}{\epsilon_a
  C_{33}^\textrm{a} +  {e_{33}^\textrm{a}}^2} ,
\end{eqnarray}
and the fully-coupled electric field by
\begin{multline}
  \label{eq:EAlGaN}
  \left . E_z^\textrm{a} \right |_{t_b \rightarrow \infty} =
  \frac{C_{33}^\textrm{a} P^\textrm{net} }{\epsilon_a
  C_{33}^\textrm{a} + {e_{33}^\textrm{a}}^2}  \\ - 2 \left (
  \frac{e_{31}^\textrm{a} C_{33}^\textrm{a} - e_{33}^\textrm{a}
  C_{13}^\textrm{a}}{\epsilon_a C_{33}^\textrm{a} +
  {e_{33}^\textrm{a}}^2} \right ) \left . \gamma_{xx}^\textrm{a}
  \right |_{t_b \rightarrow \infty} .
\end{multline}
In Eqs.\ (\ref{eq:ezza_std_hfet}), (\ref{eq:EAlGaN_std}),
(\ref{eq:ezzcoupled}) and (\ref{eq:EAlGaN}), $ \gamma_{xx}^\textrm{a}
$ takes the asymptotic limit $ ( a_b - a_a )/ a_a $ as $ t_b
\rightarrow \infty $.  The first term in Eq.\ (\ref{eq:ezzcoupled}) is
the result from the standard model.  The other two terms are due to
electromechanical coupling.

It is worth pointing out that the foregoing results also apply to a
free-standing AlGaN/GaN superlattice in which the slab of Fig.\
\ref{fig:schematic} forms a period of the superlattice.  This outcome
follows from enforcing the pseudomorphic boundary condition throughout
the entire cross-section of the slab, i.e. for all $z$.  In turn, this
result is a consequence of the simplifying assumption made at the
outset that $ \gamma_{ij} = 0 $ for $ i \ne j $.  In addition, the
electrostatic potential will be subject to periodic boundary
conditions at $ z = 0 $ and $ z = t_a + t_b $.  We can assume without
loss of generality that the potential in the superlattice case can be
set to zero at the two ends, as we have done in the present case.  A
non-zero potential at the two ends in the periodic system simply
represents a rigid shift of the potential distribution function and
does not change the electric field.  Thus a superlattice formed from a
stack of slabs discussed in the present work will have identical
strains and electric fields within each period as obtained for our
model structure.

\section{Results and discussion}
\label{sec:results}
Table \ref{tab:material} shows the material
parameters\cite{Bernardini,Wright,Yim,Maruska,Tsubouchi,Shur1,Chin}
used in the calculations.  The signs of the polarization parameters
are defined in relation to the [0001] direction: a negative sign means
that the vector is in the $ [ 000 \bar{1} ] $ direction.

\begin{figure}
  \begin{center}
    \includegraphics[width=3.2in]{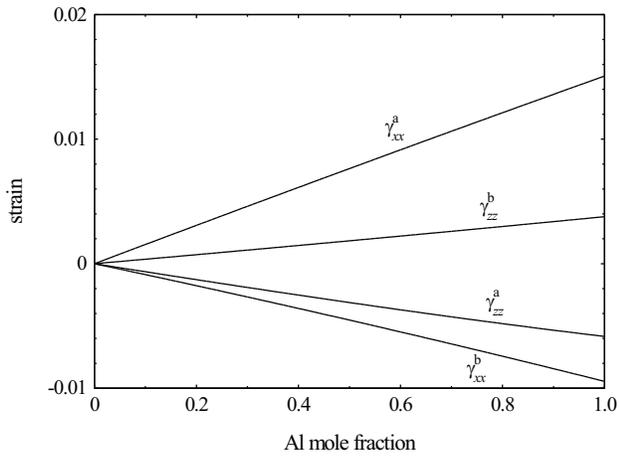}
  \end{center}
  \caption{Calculated strain for the structure of Fig.\
    \protect\ref{fig:schematic} as a function of the Al fraction with
    $ t_a $ = 300 \AA{}, $ t_b $ = 500 \AA{}, and $ n_\textrm{2D} = 0
    $, using the fully-coupled model.}
  \label{fig:freestr}
\end{figure}

For the model structure of Fig.\ \ref{fig:schematic}, we choose $ t_a
$ = 300 \AA{} and $ t_b $ = 500 \AA{}.  The 300 \AA{} AlGaN layer is
typical of most high-power HFETs.  It is recognized that the GaN layer
used in the model structure is much thinner than that permitted by
current technology.  For example, using a laser lift-off process and
growth via metal-organic chemical vapor deposition (MOCVD),
free-standing layers of nitrides can be produced successfully only for
relatively thick layers\cite{Kelly} in the vicinity of 5 $\mu$m.  The
free-standing layers are even thicker for epilayers grown by hydride
vapor phase epitaxy (HVPE), reaching between 250 -- 300
$\mu$m.\cite{Kelly2,Darak} The fully-coupled model presented herein is
quite general and, as shown in Sec.\ \ref{sec:model}, can readily
reproduce the results for thick GaN.  Figure \ref{fig:freestr} shows
the calculated strain tensor as a function of the Al fraction using
the fully-coupled model for a model structure that is depleted, i.e. $
n_\textrm{2D} = 0 $.  Following standard convention, a positive sign
indicates dilation and a negative sign contraction.  Unlike the
situation of a AlGaN layer on a semi-infinite GaN buffer, both layers
are now strained, with the in-plane strain tensile in the AlGaN layer
and compressive in the GaN layer.

\begin{figure}
  \begin{center}
    \includegraphics[width=3.2in]{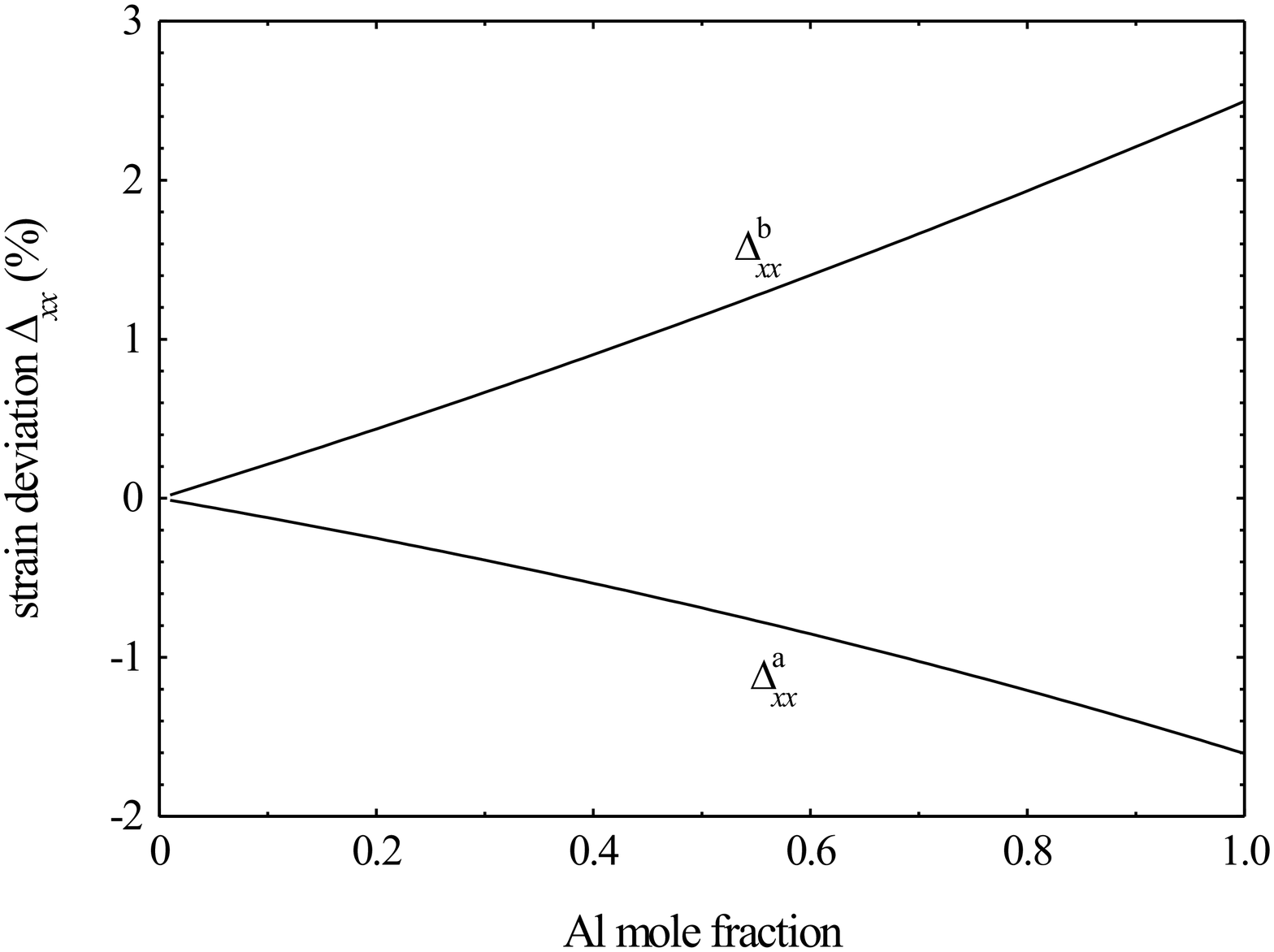}
  \end{center}
  \caption{Deviation of the uncoupled from the fully-coupled in-plane
    strain in the bi-layer slab with $ n_\textrm{2D} = 0 $.}
  \label{fig:errexx}
\end{figure}

\begin{figure}
  \begin{center}
    \includegraphics[width=3.2in]{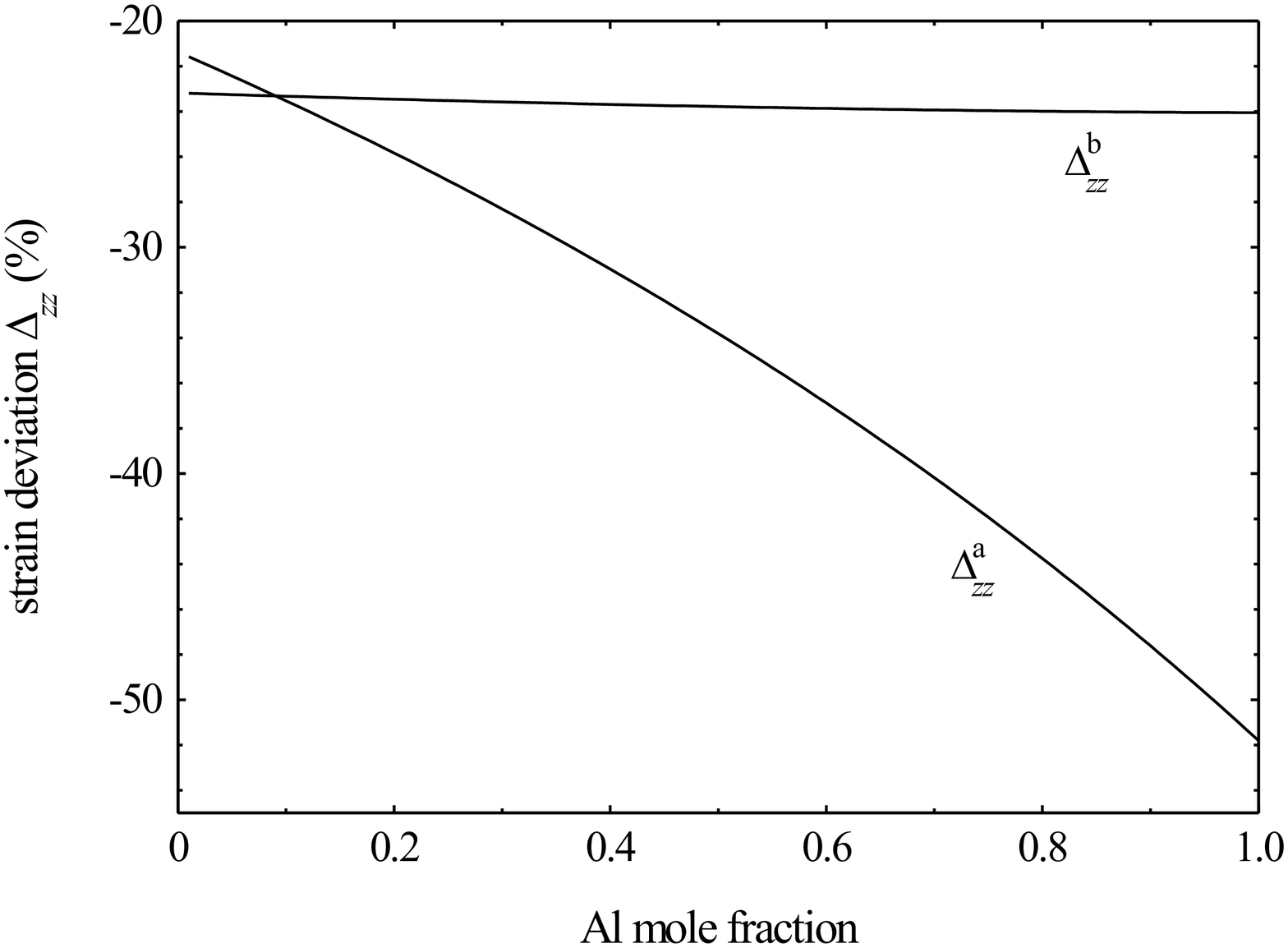}
  \end{center}
  \caption{Deviation of the uncoupled from the fully-coupled
    out-of-plane strain in the bi-layer slab with $ n_\textrm{2D} = 0
    $.}
  \label{fig:errezz}
\end{figure}

Next we examine the deviation between the fully-coupled and uncoupled
models for a depleted slab.  This deviation $ \Delta $ is defined as
\begin{equation}
  \label{eq:error}
  \Delta^\textrm{a(b)}_{ii} = \frac{\gamma_{ii}^\textrm{a(b)} -
  \gamma_{ii}^\textrm{a(b) uncoupled}}{\gamma_{ii}^\textrm{a(b)}}
\end{equation}
where $ i $ is $x$ or $z$, the superscript ``a'' or ``b'' refers to a
particular layer, and $ \gamma_{ii} $ is given by the solution to Eq.\
(\ref{eq:matrix}) for the coupled case and by Eqs.\
(\ref{eq:exxa_std}) -- (\ref{eq:ezzb_std}) for the uncoupled case.
The calculated results are shown in Figs.\ \ref{fig:errexx} and
\ref{fig:errezz} for the in-plane and out-of-plane strains,
respectively.  It is seen from Fig.\ \ref{fig:errexx} that the
deviation for $ \gamma_{xx} $ is quite small.  A possible reason is
that the electromechanical coupling of $ \gamma_{xx} $ into the
equation of state [Eq.\ (\ref{eq:state})] occurs through the
piezoelectric coefficient $ e_{31} $ which is much smaller than $
e_{33} $, as seen from Table \ref{tab:material}.  The
electromechanical coupling of $ \gamma_{zz} $ into the equation of
state is through $ e_{33} $ which is quite large, especially for AlN.
It is clear that the error in the calculated strain along the growth
axis is quite significant, reaching 30 \% in the AlGaN layer for a
mole fraction of 0.3.  This error will, in turn, have a significant
impact on the calculation of electronic and optical properties that
depend on accurate knowledge of the strain.  Some examples of such
quantities include the eigenstates and the complex dielectric function
of the slab.

\begin{figure}
  \begin{center}
    \includegraphics[width=3.2in]{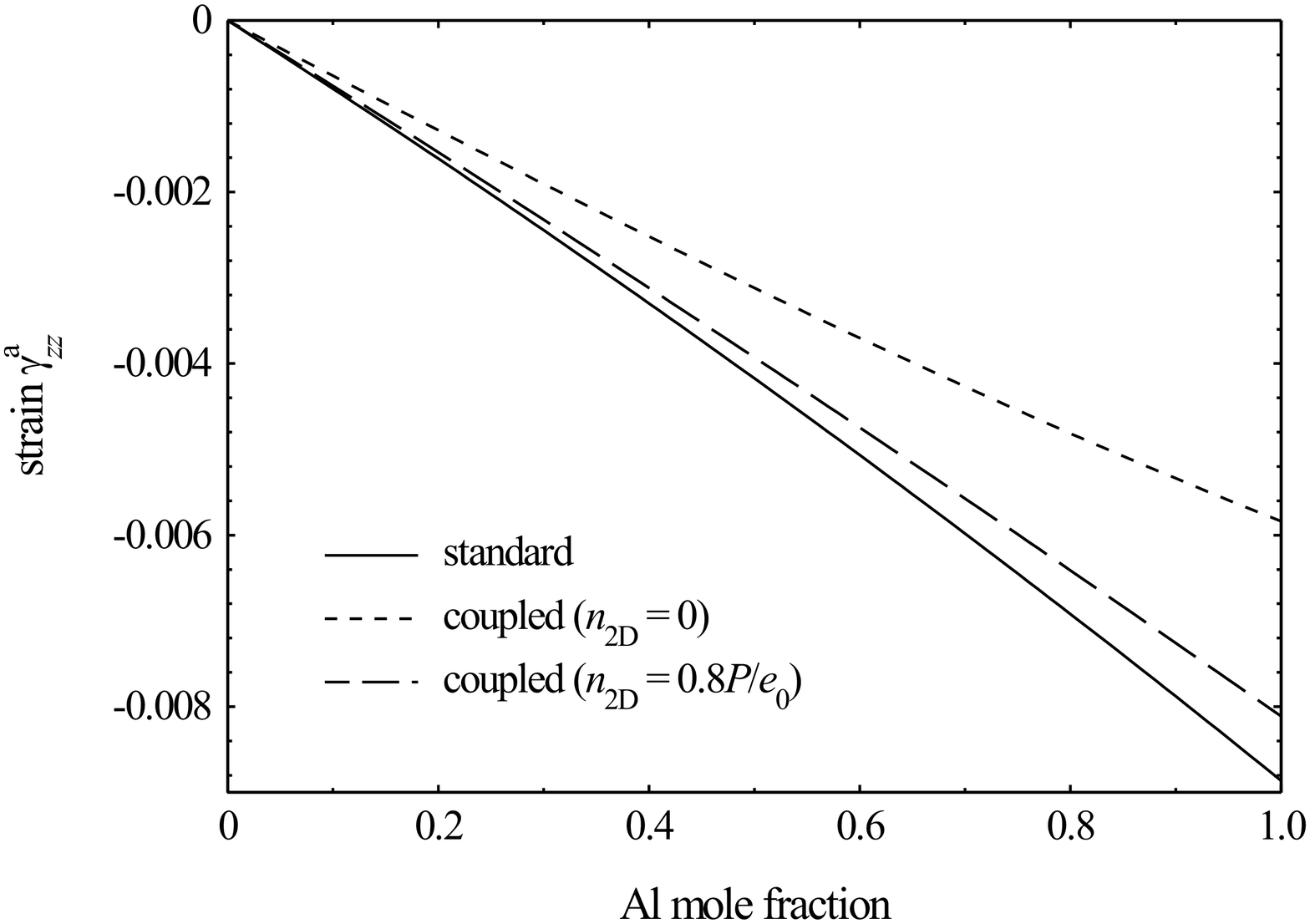}
  \end{center}
  \caption{Calculated out-of-plane strain in the AlGaN layer for the
    standard and coupled models.  Two coupled cases are shown, one
    without free electrons and one with free electrons.}
  \label{fig:ezza}
\end{figure}

\begin{figure}
  \begin{center}
    \includegraphics[width=3.2in]{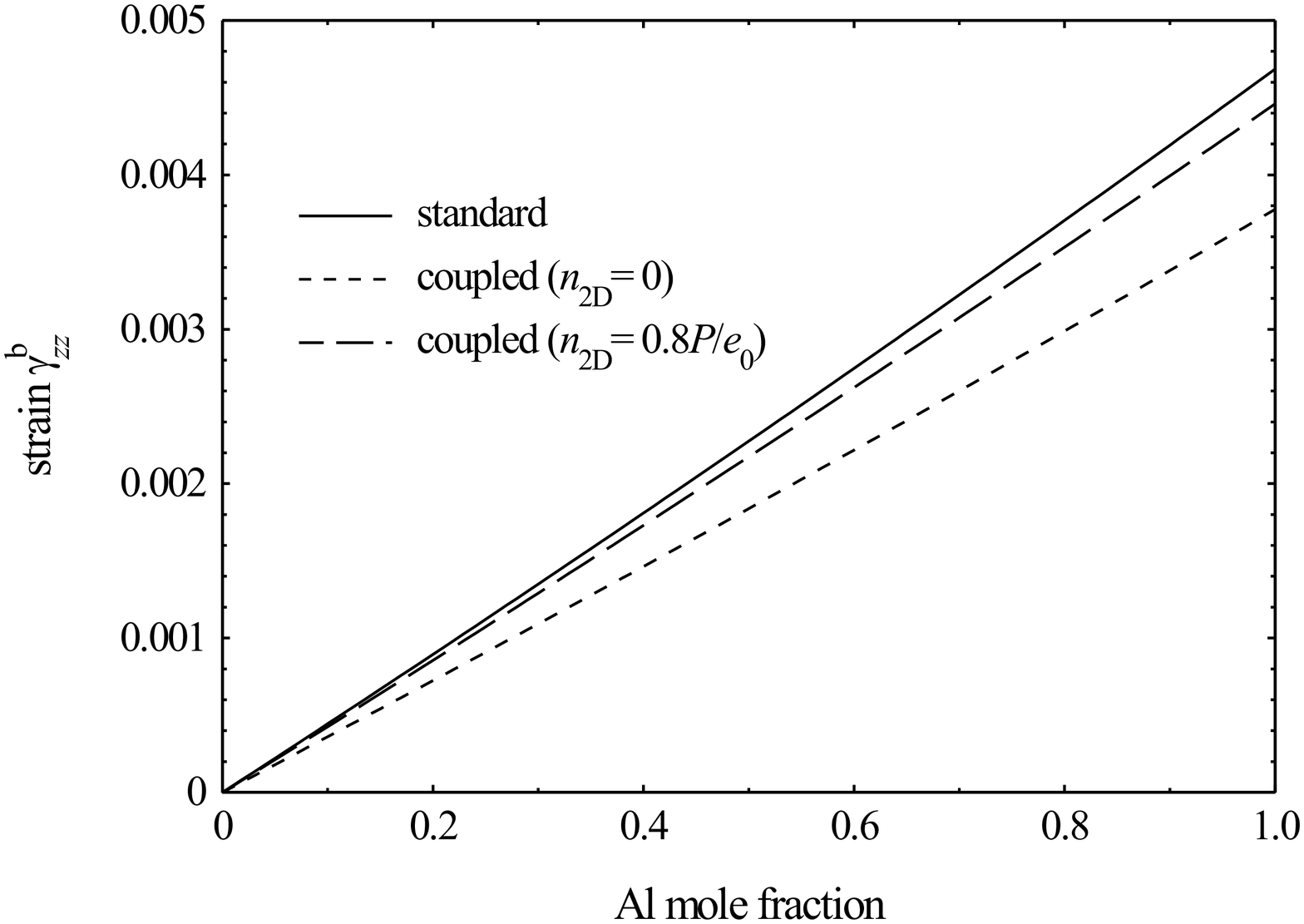}
  \end{center}
  \caption{Calculated out-of-plane strain in the GaN layer for the
    standard and coupled models.  Two coupled cases are shown, one
    without free electrons and one with free electrons.}
  \label{fig:ezzb}
\end{figure}

So far it has been assumed that the slab is depleted with the only charge
present being a fixed space charge $P$ given by
\begin{equation}
  \label{eq:spacecharge}
  P = P^\textrm{s (b) } - P^\textrm{s (a) } + P^\textrm{p (b) } -
  P^\textrm{ p (a) } ,
\end{equation}
where $ P^\textrm{p (a)} $ and $ P^\textrm{p (b)} $ are the
piezoelectric polarizations
\begin{equation}
  \label{eq:Ppa}
  P^\textrm{p (a) } = 2 e_{31}^\textrm{a} \gamma_{xx}^\textrm{a} +
  e_{33}^\textrm{a} \gamma_{xx}^\textrm{a} ,
\end{equation}
and
\begin{equation}
  \label{eq:Ppb}
  P^\textrm{p (b) } = 2 e_{31}^\textrm{b} \gamma_{xx}^\textrm{b} +
  e_{33}^\textrm{b} \gamma_{xx}^\textrm{b} .
\end{equation}
For a cation-faced structure, the standard growth orientation for most
HFETs, $P$ is positive.\cite{Ambacher1} Depending on the growth and
surface conditions, donors may be present and may contribute free
electrons to the interface.  Under certain conditions, the space
charge may be partially screened by a 2DEG.  For instance, using our
own self-consistent Schr\"odinger-Poisson model,\cite{hfetcharge} we
find that the 2DEG is almost 90 \% of the magnitude of the fixed space
charge.  We assume, conservatively, that $ n_\textrm{2D} = 0.8 P / e_0
$ and examine the effect of free-electron screening on the
electromechanical coupling.  Figures \ref{fig:ezza} and \ref{fig:ezzb}
show $ \gamma_{zz}^\textrm{a} $ and $ \gamma_{zz}^\textrm{b} $,
respectively, calculated using the standard and coupled models.  The
in-plane strains are not shown, since the effect of electromechanical
coupling on the in-plane strain components is weak.  It is evident
from the calculated strains that the effect of coupling is to reduce
the magnitude of the strains relative to those of the standard model,
especially when the structure is depleted.  This result is not
surprising, since we would expect the piezoelectric fields to induce
forces to oppose any applied forces such as those due to the
pseudomorphic interface condition.  Another result that is evident
from Figs.\ \ref{fig:ezza} and \ref{fig:ezzb} is that the 2DEG
screening brings the strains of the coupled model closer to those of
the standard model.  As will be seen shortly, the effect of screening
is to reduce the built-in electric fields.  This, in turn, will reduce
the electromechanical coupling, as seen from Eq.\ (\ref{eq:state}).

\begin{figure}
  \begin{center}
    \includegraphics[width=3.2in]{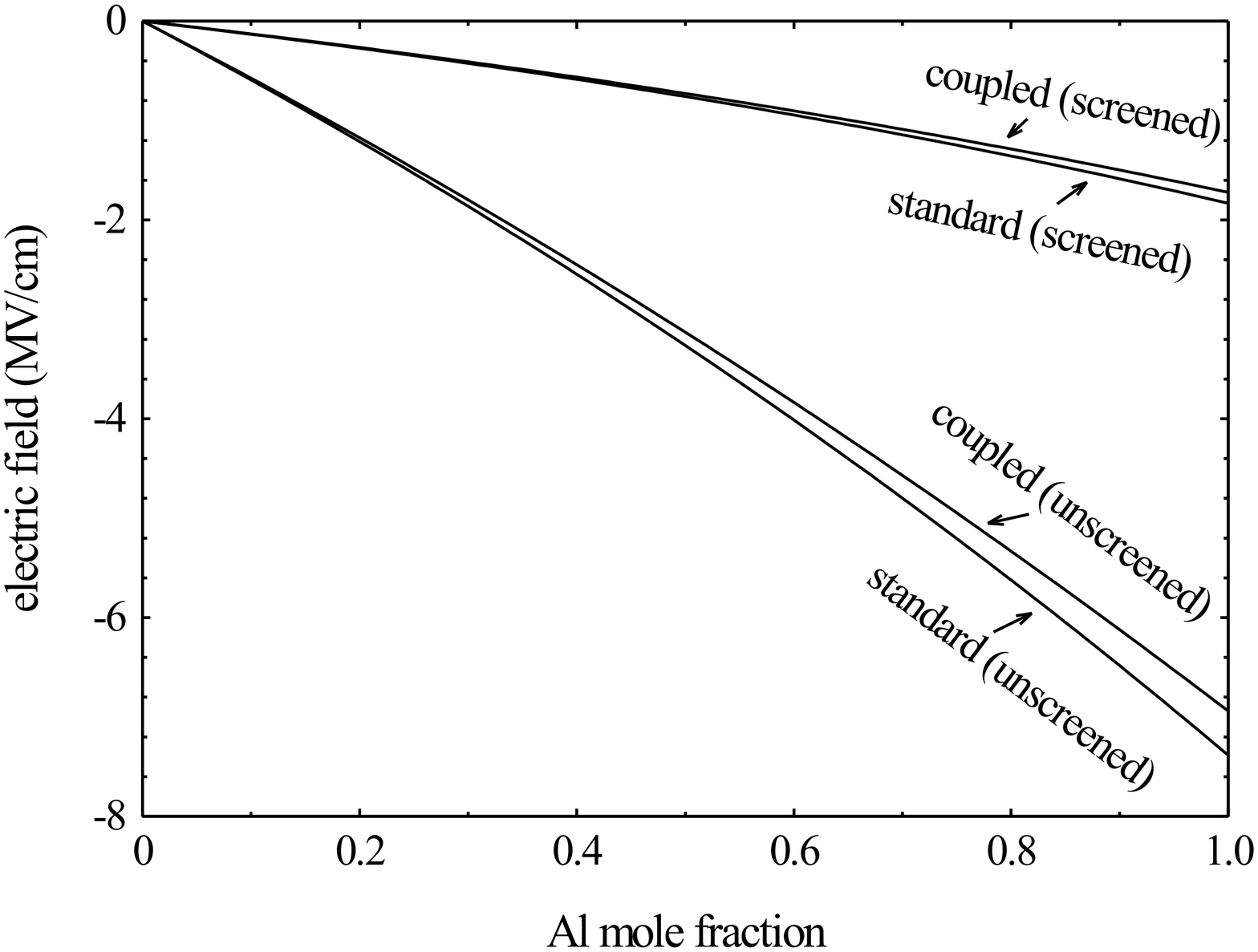}
  \end{center}
  \caption{Calculated electric field in the AlGaN layer for the
    standard and fully-coupled models showing the effect of
    free-electron screening in each case.}
  \label{fig:Eza}
\end{figure}

\begin{figure}
  \begin{center}
    \includegraphics[width=3.2in]{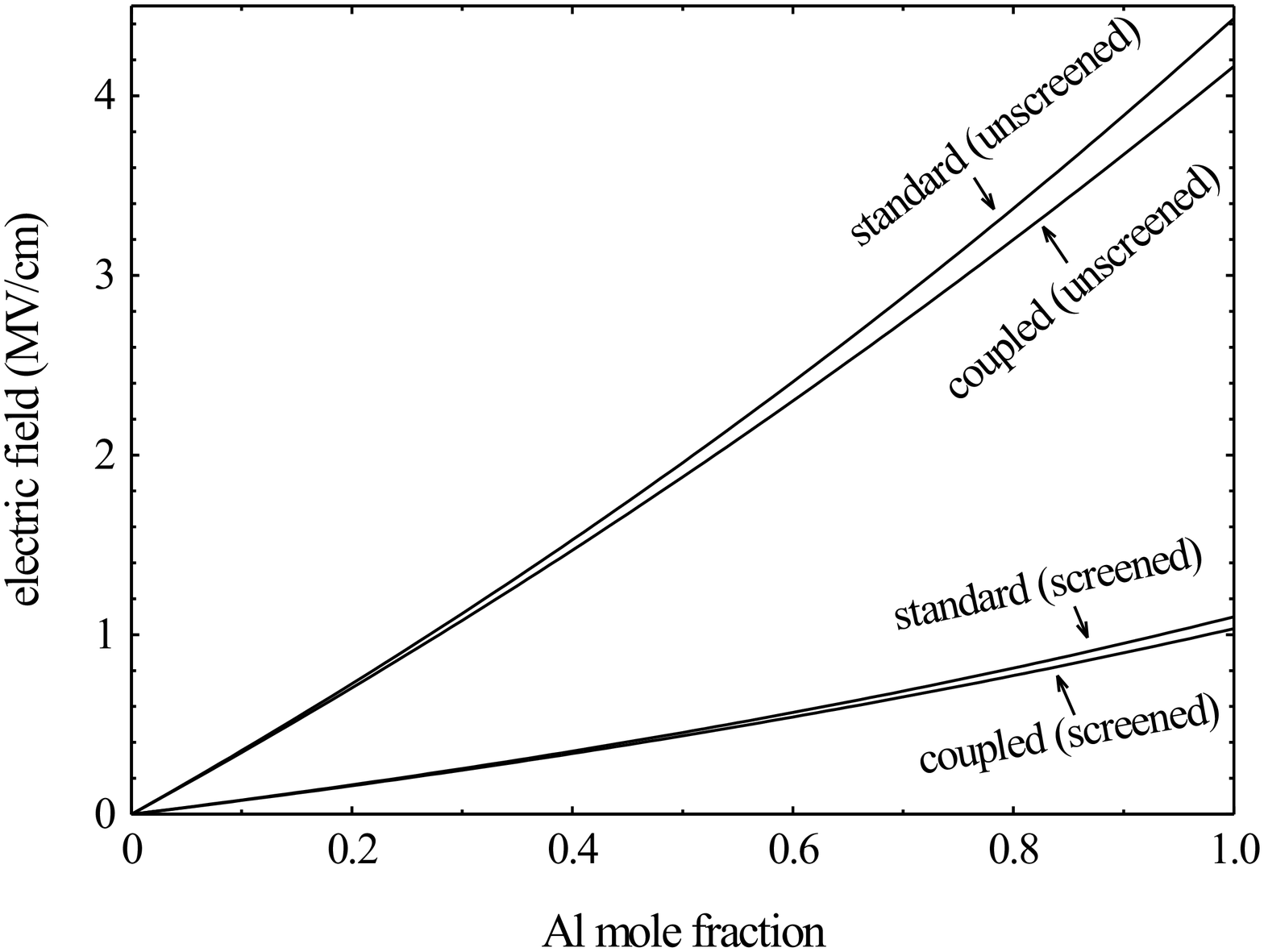}
  \end{center}
  \caption{Calculated electric field in the GaN layer for the
    standard and fully-coupled models showing the effect of
    free-electron screening in each case.}
  \label{fig:Ezb}
\end{figure}

Figures \ref{fig:Eza} and \ref{fig:Ezb} show the calculated electric
fields in the AlGaN and GaN layers, respectively, using the standard
and coupled models.  These fields help to explain, in part, the large
deviation of the standard strain fields from their coupled
counterparts.  As an example, for a Al fraction of 0.3, the electric
field in the AlGaN layer is about 2 MV/cm, provided that the structure
is depleted.  Such a large field, if present, will give rise to a
strong electromechanical coupling and, therefore, a large discrepancy
between the standard and coupled models.  When there is a 2DEG,
however, the large space charge that gives rise to the electric field
is partially neutralized, depending on the magnitude of the 2DEG.  For
$ n_\textrm{2D} = 0.8 P $, the field in the AlGaN layer is reduced to
about 0.4 MV/cm for a Al fraction of 0.3.  Evidence for the smaller
field is seen in recent photoreflectance (PR)
measurements\cite{Goldhahn,Shokh} on AlGaN/GaN heterostructures in
which barrier fields in the vicinity of 0.3 MV/cm were reported for
250 \AA{} thick barriers with Al fractions in the range 0.06 -- 0.1.
An electric field of opposite sign occurs in the GaN layer with
similar trends between the coupled and standard models.  Its magnitude
is decreased with the thickness of the GaN layer.  It is recognized
that the calculation of the electric field in the GaN layer is more
involved  than the classical model presented here allows.  The
formation of the notch in the conduction band edge in which the 2DEG
resides is reproducible only within the scope of a quantum-mechanical
calculation.  In spite of approximations used, the calculated fields
shown here appear to be quite plausible.

\section{Summary and Conclusions}
\label{sec:summary} In summary, a fully-coupled electromechanical
model has been presented for the strain and electric fields in a
free-standing bi-layer AlGaN/GaN slab.  The electric enthalpy of
the slab, with contributions from the piezoelectric and
spontaneous polarizations, is minimized subject to the constraints
imposed by the pseudomorphic interface condition and the Poisson
equation. Closed-form expressions for the strain components and
electric field in each layer are obtained. The results for the
bi-layer system are also appropriate for freestanding binary
superlattices with negligible bowing. Large discrepancies between
the standard and coupled models are found for depleted structures
due to large built-in electric fields. The electric fields are
reduced when the polarization-induced space charge is screened by
a 2DEG. As a consequence, the discrepancy between the standard and
coupled models is reduced when a 2DEG is present.

\acknowledgments The work of BJ was partially supported by the Air
Force Office of Scientific Research (AFOSR) and performed at Air Force
Research Laboratory, Materials and Manufacturing Directorate
(AFRL/MLP), Wright Patterson Air Force Base under USAF Contract No.\
F33615-00-C-5402.

\appendix
\section{Matrix solution of the strain field}
\label{app:matrix}
The matrix elements required to solve Eq.\ (\ref{eq:matrix}) are given
in this appendix.  As before, ``a'' refers to the AlGaN layer and
``b'' to the GaN layer.
\begin{multline}
  \label{eq:a11}
  A_{11} =  t_a a_b^2 ( C_{11}^\textrm{a} +  C_{12}^\textrm{a}) (
  t_a \epsilon_a + t_b \epsilon_b ) \\ +  t_b  a_a^2 ( t_a
  \epsilon_b + t_b \epsilon_a ) (
  C_{11}^\textrm{b} + C_{12}^\textrm{b} ) \\ +
  2 t_a t_b ( e_{31}^\textrm{b} a_a - e_{31}^\textrm{a} a_b ) ^2 ,
\end{multline}
\begin{multline}
  \label{eq:a12}
  A_{12} = t_a a_b^2 C_{13}^\textrm{a} ( t_a \epsilon_b + t_b
  \epsilon_a ) \\ + t_a t_b a_b e_{33}^\textrm{a}
  ( e_{31}^\textrm{a} a_b - e_{31}^\textrm{b} a_a )  ,
\end{multline}
\begin{multline}
  \label{eq:a13}
  A_{13} = t_b a_a a_b C_{13}^\textrm{b} ( t_a \epsilon_b + t_b
   \epsilon_a ) \\ - t_a t_b a_b e_{33}^\textrm{b} ( e_{31}^\textrm{a}
   a_b - e_{31}^\textrm{b} a_a ) ,
\end{multline}
\begin{multline}
\label{eq:a21}
  A_{21} = 2 t_a a_b C_{13}^\textrm{a} ( t_a \epsilon_b + t_b
  \epsilon_a ) \\ + 2 t_a t_b e_{33}^\textrm{a} ( e_{31}^\textrm{a} a_b -
  e_{31}^\textrm{b} a_a ) ,
\end{multline}
\begin{equation}
  \label{eq:a22}
  A_{22} = t_a a_b C_{33}^\textrm{a} ( t_a \epsilon_b + t_b \epsilon_a
) + t_a t_b a_b {e_{33}^\textrm{a}}^2 ,
\end{equation}
\begin{equation}
  \label{eq:a23}
  A_{23} = - t_a t_b a_b e_{33}^\textrm{a} e_{33}^\textrm{b} ,
\end{equation}
\begin{multline}
  \label{eq:a31}
  A_{31} = 2 t_a t_b e_{33}^\textrm{b} ( e_{31}^\textrm{b} a_a -
  e_{31}^\textrm{a} a_b ) \\ + 2 t_b a_a C_{13}^\textrm{b} ( t_a
  \epsilon_b + t_b \epsilon_a ) ,
\end{multline}
\begin{equation}
  \label{eq:a32}
  A_{32} =   - t_a t_b a_b e_{33}^\textrm{a} e_{33}^\textrm{b} ,
\end{equation}
\begin{equation}
  \label{eq:a33}
  A_{33} = t_b a_b C_{33}^\textrm{b} ( t_a \epsilon_b + t_b \epsilon_a
  ) + t_a t_b a_b { e_{33}^\textrm{b}}^{2} ,
\end{equation}
\begin{multline}
  \label{eq:b1}
  B_1 = t_a t_b ( e_{31}^\textrm{a} a_b - e_{31}^\textrm{b} a_a )
  \\ \times
  \left [ a_b ( P^\textrm{s (b)} - P^\textrm{s (a)} + e_0
  n_\textrm{2D} )  +
    2 e_{31}^\textrm{b} ( a_a - a_b ) \right ] \\ +
   a_a
  t_b ( a_b -
  a_a ) ( C_{11}^\textrm{b} + C_{12}^\textrm{b} ) ( t_a \epsilon_b +
  t_b \epsilon_a ) ,
\end{multline}
\begin{multline}
  \label{eq:b2}
  B_2 =  2 t_a t_b e_{31}^\textrm{b} e_{33}^\textrm{a} ( a_a - a_b ) \\ +
   t_a t_b a_b e_{33}^\textrm{a} ( P^\textrm{s(b)} - P^\textrm{s(a)} +
    e_0  n_\textrm{2D} ) ,
\end{multline}
and
\begin{multline}
  \label{eq:b3}
  B_3 = - t_a t_b a_b e_{33}^\textrm{b} ( P^\textrm{s(b)} -
  P^\textrm{s(a)} + e_0 n_{2D} ) \\ + 2 ( a_b - a_a ) [ t_a
  t_b e_{31}^\textrm{b} e_{33}^\textrm{b} + t_b C_{13}^\textrm{b} (
  t_a \epsilon_b + t_b \epsilon_a ) ] .
\end{multline}

\bibliography{electrofreestand}

\begin{table}
\caption{Strain-related material parameters used in the present model.
  The elastic stiffness constants are in units of GPa and the
  piezoelectric coefficients and spontaneous polarization in units
  of C/$\textrm{m}^2$.}
  \label{tab:material}
\begin{tabular}{lccccccccc}
Material & $C_{11}$ & $C_{12}$ & $C_{13}$ & $C_{33}$ & $a$ (\AA{}) &
$ e_{31} $ & $ e_{33} $ & $ P^\textrm{s} $ & $ \epsilon / \epsilon_0 $ \\
\hline AlN & 396\tablenotemark[1] & 137\tablenotemark[1] &
108\tablenotemark[1] & 373\tablenotemark[1] &
3.112\tablenotemark[2] & $-$0.58\tablenotemark[4] &
1.55\tablenotemark[4] & $-$0.081\tablenotemark[6] &
8.5\tablenotemark[7] \\
GaN & 367\tablenotemark[1] & 135\tablenotemark[1] & 103\tablenotemark[1] &
405\tablenotemark[1] &
3.189\tablenotemark[3] & $-$0.36\tablenotemark[5] &
1\tablenotemark[5] & $-$0.029\tablenotemark[6] & 10\tablenotemark[7] \\
\end{tabular}
\tablenotetext[1]{Reference \onlinecite{Wright}.}
\tablenotetext[2]{Reference \onlinecite{Yim}.}
\tablenotetext[3]{Reference \onlinecite{Maruska}.}
\tablenotetext[4]{Reference \onlinecite{Tsubouchi}.}
\tablenotetext[5]{Reference \onlinecite{Shur1}.}
\tablenotetext[6]{Reference \onlinecite{Bernardini}.}
\tablenotetext[7]{Reference \onlinecite{Chin}.}
\end{table}

\newpage

\end{document}